\documentclass[12pt,superscriptaddress]{iopart}

\usepackage{iopams}

\usepackage{graphicx}
\usepackage{mathtext}

\begin{document}

\title[Localization of convective currents under parametric disorder]{Localization and advectional spreading of convective currents under parametric disorder}

\author{Denis S Goldobin}
\address{Institute of Continuous Media Mechanics, UB RAS,
         1 Acad.\ Korolev street, Perm 614013, Russia}
\address{Department of Mathematics, University of Leicester,
         University Road, Leicester LE1 7RH, UK}
\ead{Denis.Goldobin@gmail.com}
\author{Elizaveta V Shklyaeva}
\address{Department of Theoretical Physics, Perm State University,
         15 Bukireva street, Perm 614990, Russia}
\ead{shklyaeva-liza@yandex.ru}

\begin{abstract}
We address a problem which is mathematically reminiscent of the one of Anderson localization, although it is related to a strongly dissipative dynamics. Specifically, we study thermal convection in a horizontal porous layer heated from below in the presence of a parametric disorder; physical parameters of the layer are time-independent and randomly inhomogeneous in one of the horizontal directions. Under such a frozen parametric disorder, spatially localized flow patterns appear. We focus our study on their localization properties and the effect of an imposed advection along the layer on these properties. Our interpretation of the results of the linear theory is underpinned by numerical simulation for the nonlinear problem. Weak advection is found to lead to an upstream delocalization of localized current patterns. Due to this delocalization, the transition from a set of localized patterns to an almost everywhere intense ``global'' flow can be observed under conditions where the disorder-free system would be not far below the instability threshold. The results presented are derived for a physical system which is mathematically described by a modified Kuramoto--Sivashinsky equation and therefore they are expected to be relevant for a broad variety of dissipative media where pattern selection occurs.
\end{abstract}


\vspace{2pc}
\submitto{\it J.\ Stat.\ Mech.: Theory Exp.}

\maketitle

\section{Introduction}
The interest to the effect of localization in spatially extended systems under parametric disorder was first initiated in the context of phenomena in quantum systems; this effect, Anderson localization~\cite{Anderson-1958} (AL), has been originally discussed for such processes as spin diffusion or electron propagation in a disordered potential. Later on, AL was studied in various branches of physics in the context of wave propagation in randomly inhomogeneous acoustic~\cite{Maynard-2001}, optical media~\cite{Klyatskin-2005}, {\it etc.} The common feature for the listed works is that they deal with conservative systems (or media) in contrast to dissipative systems like in the problems of thermal convection. Paradigmatic AL is a substantially linear phenomenon which is affected by nonlinearity, up to destruction of localization, whereas dissipative systems are essentially nonlinear (without nonlinearity patterns either grow or decay exponentially in time).

In~\cite{Hammele-Schuler-Zimmermann-2006} the effect of parametric disorder on the excitation threshold in an active medium, 1D Ginzburg--Landau equation, has been discussed, but the localization effects were beyond the study scope. The localization properties of solutions to the linearized Ginzburg--Landau equation are identical to the one for the Schr\"odinger equation (where AL was comprehensively discussed; for instance,
see~\cite{Froehlich-Spencer-1984,Lifshitz-Gredeskul-Pastur-1988,Gredeskul-Kivshar-1992}), while for thermal convection it is so only under specific conditions. To the authors' knowledge, localization (or, in an alternative formulation more suitable for our problem, {\it excitation of localized modes}) in the presence of a frozen parametric disorder in fluid dynamics problems (such as thermal convection) has not been studied in the literature.

The paper is organized as follows. In \sref{sec2}, we introduce the specific physical system we deal with and corresponding mathematical model, illustrate the appearance of localized patterns, and introduce quantifiers of localization. In \sref{sec3}, the spatial density of the excitation centers of localized patterns is calculated. In \sref{sec4}, we present the derivation of localization properties from the linear theory and discus the effect of an imposed longitudinal advection on these properties. In \sref{sec5}, we underpin the results of the linear theory with numerical simulation of the non-linearized equations. In \sref{sec6}, the localization phenomenon in dissipative systems is compared to the Anderson localization in conservative systems. \Sref{concl} concludes the paper with summary of results and argumentation in support of generality of our findings.

\section{Localized patterns in large-scale thermal convection in a horizontal layer}\label{sec2}
\subsection{Physical problem formulation}\label{sec21}
\begin{figure}[!t]
\center{
  \includegraphics[width=0.52\textwidth]%
 {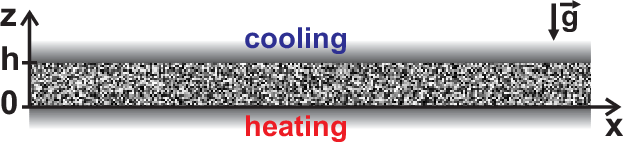}
}
  \caption{Sketch of the system and coordinate frame}
  \label{fig1}
\end{figure}

As a specific physical system, we consider a thin porous layer saturated with a fluid. The layer is confined between two impermeable horizontal plates and heated from below (Fig.\,\ref{fig1}). The coordinate frame is such that the $(x,y)$-plane is horizontal, $z=0$ and $z=h$ are the lower and upper boundaries, respectively. The bounding plates are nearly thermally insulating (in comparison to the layer) which results in a fixed heat flux through the layer boundaries. This flux $Q(x,y)$ is time-independent and inhomogeneous in space;
\[
Q(x,y)=Q_\mathrm{cr}(1+\varepsilon^2q(x,y))\,,
\]
here $Q_\mathrm{cr}$ is the threshold value of the heat flux for the case of spatially uniform heating (convective flows are excited above the threshold), $\varepsilon^2$ is a characteristic magnitude of local relative deviations of the heat flux from the critical value $Q_\mathrm{cr}$. In this paper, we restrict our consideration to the case of $q=q(x)$ and the flows homogeneous along the $y$-direction. Also, we admit a pumping of the fluid along the layer.

For nearly thermally insulating boundaries and a uniform heating, the marginal convective instability of the layer is long-wavelength~\cite{Sparrow_etal-1964,Ingham-Pop-1998} (in other words, large-scale), which means that the horizontal scale $L$ of the flow is large compared to the layer height $h$, $L\sim\varepsilon^{-1}h$. At nearly critical conditions (close to the instability threshold of the uniformly heated system), convective currents are still long-wavelength. For large-scale convection the temperature perturbations $\theta=\theta(x)$ are almost uniform along $z$ and, according to~\cite{Goldobin-Shklyaeva-BR-2008}, the system is governed by the dimensionless equation
\begin{equation}
\dot{\theta}=\left(-u\theta-\theta_{xxx}-q(x)\theta_x+(\theta_x)^3\right)_x,
\label{eq2-01}
\end{equation}
where $u$ is the $x$-component of the imposed advection (through-flow) velocity. Noteworthy, the same equation governs some other fluid dynamical systems~\cite{Knobloch-1990,Shtilman-Sivashinsky-1991,Aristov-Frick-1989}. For equation~\eref{eq2-01} the length scale is set to $L=\sqrt{21/2}\,h/\varepsilon$ (the geometric factor $\sqrt{21/2}$ is needed to set all the coefficients of the equation to $1$) and thus the dimensionless layer height $h=\sqrt{21/2}\,\varepsilon\ll 1$.

In equation~\eref{eq2-01}, the term $(-q(x)\theta_x)_x$ effectively functions as heat diffusion with inhomogeneous heat diffusion coefficient $q(x)$ which can be positive, leading to decay of temperature inhomogeneities, or negative, leading to inhomogenization. The term $(-\theta_{xxxx})$ suppresses short waves and prevents the formation of field discontinuities and cusps when $q(x)$ is negative. The nonlinear term $((\theta_x)^3)_x$ bounds the growth of patterns $\theta(x)$ when they become finite-amplitude.

Frozen-disorder-induced effects in the system can be observed also when the heating is uniform but macroscopic properties of the porous matrix are weakly inhomogeneous (inhomogeneity of porosity, permeability, and heat conductivity is inevitable in real systems). Moreover, the system will be governed by the same equation~\eref{eq2-01}, with the only difference in the relationship between $q(x)$ and physical parameter inhomogeneities~\cite{Goldobin-Shklyaeva-BR-2008}. Nonetheless, we consider an inhomogeneous heating in order to make it more obvious that our findings can be observed for convection without a porous matrix as well.

Although equation~\eref{eq2-01} is derived for a large-scale inhomogeneity, {\it i.e.},
 $h|q_x|/|q|\ll 1$,
one may set such a hierarchy of small parameters, namely $\varepsilon\ll(h|q_x|/|q|)^2\ll1$, that, on the one hand, the long-wavelength approximation remains valid and, on the other hand, the frozen inhomogeneity may be represented by a $\delta$-correlated Gaussian noise:
\[
q(x)=q_0+\xi(x),\; \left\langle\xi(x)\right\rangle=0,\;
\left\langle\xi(x)\xi(x')\right\rangle=2D\delta(x\!-\!x'),
\]
where $q_0$ is the mean deviation from the instability threshold of the system without disorder (hereafter shortly referred as ``mean supercriticality'').

The noise strength $D$ may be set to $1$ by means of the following rescaling of variables and parameters: $(x,t,q)\to(D^{-1/3}x,D^{-4/3}t,D^{2/3}q)$. Let us see this explicitly. Performing the rescaling, one should keep in mind, that the statistical effect of a $\delta$-correlated noise is determined by the integral of its correlation function which is influenced by coordinate stretching. In detail, under the rescaling $q=\mathcal{Q}\tilde{q}$ (equivalently,
$\xi=\mathcal{Q}\tilde{\xi}$), $x=\mathcal{X}\tilde{x}$, and
$t=\mathcal{T}\tilde{t}$, one finds
$\langle\xi(x')\,\xi(x'+x)\rangle=2D\,\delta(x)
=2D\,\mathcal{X}^{-1}\delta(\tilde{x})
=\mathcal{Q}^2\langle\tilde{\xi}(\tilde{x}')\,\tilde{\xi}(\tilde{x}'+\tilde{x})\rangle$.
In order to operate with a normalized noise, {\it i.e.}, $\langle\tilde{\xi}(\tilde{x}')\,\tilde{\xi}(\tilde{x}'+\tilde{x})\rangle=2\delta(\tilde{x})$,
one should choose $\mathcal{Q}^2=D/\mathcal{X}$; to preserve equation~\eref{eq2-01} unchanged, one has to claim
$\mathcal{T}=\mathcal{X}^4$ and $\mathcal{Q}=\mathcal{X}^{-2}$.
These conditions yield $\mathcal{X}=D^{-1/3}$,
$\mathcal{T}=D^{-4/3}$, and $\mathcal{Q}=D^{2/3}$. Henceforth, we set $D=1$. As the layer is practically not infinite, one should be subtle with the small-noise limit which implies a shrinking of the dimensionless length of the system owing to the rescaling of the spatial variable.

In specific physical systems described by equation~\eref{eq2-01}, properties of various physical fields may differ, and, simultaneously, for particular physical phenomena in these systems, properties of one or another field may play a decisive role. For instance, transport of a pollutant~\cite{Goldobin-Shklyaeva-2009-Goldobin-2010} is determined by the velocity field which not necessarily possesses the same localization properties as the temperature field. Therefore, relations between various fields and field $\theta$ are worth consideration, even though all the major findings of the paper can be provided in terms of $\theta$ regardless to a specific origin of equation~\eref{eq2-01}. As derived in~\cite{Goldobin-Shklyaeva-BR-2008} for a porous layer we consider, the fluid velocity field of the convective flow is
\begin{equation}
\vec{v}=\frac{\partial\Psi}{\partial z}\vec{e}_x
 -\frac{\partial\Psi}{\partial x}\vec{e}_z\,,
 \qquad
\Psi=f(z)\,\psi(x,t)\,,
\label{flow}
\end{equation}
where $\psi(x,t)\equiv\theta_x(x,t)$ is the stream function amplitude and $f(z)=3\sqrt{35}D^{-1}h^{-3}z(h-z)$. It is also the result of work~\cite{Goldobin-Shklyaeva-BR-2008}, that the contribution of imposed advection $u$ should not be presented here due to its smallness in comparison to $v$ ($u\sim 1$ {\it vs.}\ $v\sim h^{-2}$). In spite of its smallness, the advection $u$ influences the system dynamics due to its spatial properties; $u$ provides a fluid gross flux through a cross-section of the layer, $\int_0^hu\,dz=uz\ne 0$, whereas for the convective flow $\vec{v}$ this gross flux
 $\int_0^hv^{x}dz
 =\int_0^h(\partial\Psi/\partial z)dz
 =\Psi(z=h)-\Psi(z=0)=0$
(cf.\ equation~\eref{flow}). When the gross flux of a certain flow is zero, the transport (such as heat transfer) by this flow is essentially less efficient compared to the case of a nonzero gross flux.


\begin{figure}[!t]
\center{
  \includegraphics[width=0.70\textwidth]%
 {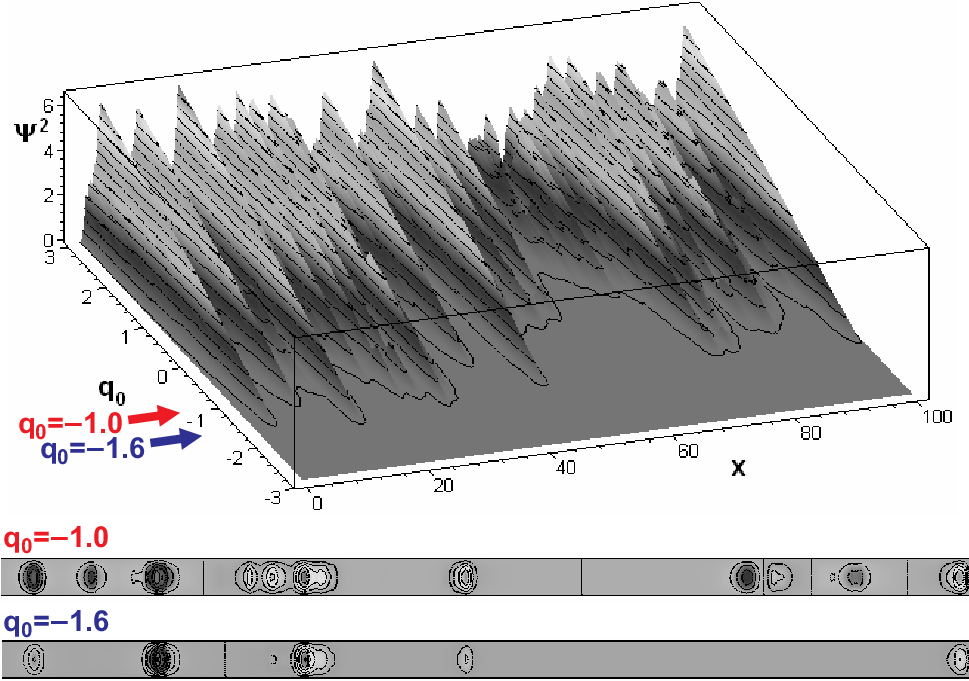}
}
  \caption{The upper figure plots the squared stream function amplitude $\psi^2(x,q_0)$ for the steady time-independent solution of equation~\eref{eq2-01} as a function of $q_0$ for a sample $\xi(x)$ and $u=0$. The lower figures present flow stream lines for $q_0=-1.0$ and $-1.6$ (the scales of $x$ and $z$ are different).
  }
\label{fig2}
\end{figure}

\subsection{Localized patterns}
\label{sec22}
In \fref{fig2}, one can see the demonstration of patterns arising in the system~\eref{eq2-01}.  For $q_0>0$, where convective currents would arise even without disorder, a nearly everywhere-intense irregular flow is excited. For $q_0<0$, where convective currents and temperature perturbations would decay in the absence of disorder, convective flow is less intense; as $q_0$ decreases, the flow turns into a set of localized current patterns, which disappear gradually. The larger the magnitude of negative $q_0$, the smaller is the spatial density of the current patterns. Where the patterns are distinguishably isolated from each other, the issue of their localization properties becomes relevant. These localized patterns and their localization properties are in the focus of this paper.

When some pattern is localized near the point $x_0$, at the distance from $x_0$,
\[
\theta_x\propto\left\{
\begin{array}{cc}
\exp(-\gamma_-|x-x_0|),&(x-x_0)u>0;\\[3pt]
\exp(-\gamma_+|x-x_0|),&(x-x_0)u<0
\end{array}\right.
\]
(for example, see localized patterns in figures~\ref{fig6}a and \ref{fig7}a). Here $\gamma_-$ and $\gamma_+$ are the down- and upstream localization exponents, respectively; the inverse value $\lambda_\pm=\gamma_\pm^{-1}$ is referred to as the ``localization length''.

With numerical simulation we found only the time-independent solutions to be stable steady states in the dynamical system~\eref{eq2-01} for small enough $u$. Therefore, the primary objects of our interest are time-independent solutions and their localization properties. In the region of exponentially decaying tails the solutions are small and thus the exponents may be found from the linearized form of equation~\eref{eq2-01}, which can be one time integrated with respect to $x$ in the time-independent case;
\begin{equation}
u\theta+\theta'''+[q_0+\xi(x)]\theta'=const\equiv S\,
\label{eq2-02}
\end{equation}
(the prime denotes the $x$-derivative). For $u\ne 0$ the substitution $\theta\to\theta+S/u$ turns integration constant $S$ to zero.

For $u=0$ ($S=0$), $\theta'$ is governed by the stationary Schr\"odinger equation with the mean supercriticality $q_0$ instead of the state energy $E$ and $(-\xi(x))$ instead of the potential $U(x)$;
\[
q_0\theta'=-\frac{\rmd^2}{\rmd x^2}\theta'+(-\xi(x))\theta'.
\]
For the Schr\"odinger equation with a $\delta$-correlated potential---a paradigmatic model for the AL---the localization is comprehensively studied; all states (for any energy) are localized ({\it e.g.}, \cite{Froehlich-Spencer-1984,Lifshitz-Gredeskul-Pastur-1988,Gredeskul-Kivshar-1992}). While the localization exponents $\gamma_\pm$ for this case are reported in the literature, the spatial density of excited localized modes in the system~\eref{eq2-01} is not known even for $u=0$. The phenomenon of localization for $u\ne0$ was not studied as well.

\section{Spatial density of localized patterns for $u=0$}\label{sec3}
Let us estimate the likelihood of the occurrence of a non-decaying pattern near certain point $x_0$ assuming that all other excited patterns are distant enough from $x_0$. One can multiply equation~\eref{eq2-01} (with $u=0$) by $\theta(x)$ and integrate over space domain $[x_0-H,x_0+H]$ which does not contain any other patterns, but is large enough to assume $\theta(x)\approx0$ at its edges.  After partial integration one finds
\[
\frac{\rmd}{\rmd t}\int\limits_{x_0-H}^{x_0+H}\frac{\theta^2}{2}\rmd x
 =-\int\limits_{x_0-H}^{x_0+H}(\theta_{xx})^2\rmd x
 +\int\limits_{x_0-H}^{x_0+H}q(x)(\theta_x)^2\rmd x
 -\int\limits_{x_0-H}^{x_0+H}(\theta_x)^4\rmd x\,.
\]
All integrals here except for the convolution of $(\theta_x)^2$ with $q(x)$ are strictly non-negative. Hence, the condition for a small perturbation $\theta(x)$ to be non-decaying is the inequality
\begin{equation}
-\int(\theta_{xx})^2\rmd x+\int q(x)(\theta_x)^2\rmd x>0\,.
\label{ineq01}
\end{equation}

With decomposition
 $\int q(x)(\theta_x)^2\rmd x=q_0\int(\theta_x)^2\rmd x+\int\xi(x)(\theta_x)^2\rmd x$,
one can see, that for negative $q_0$, the only term which can make a non-negative contribution to the l.h.s.\ part of inequality~\eref{ineq01} is $\int\xi(x)(\theta_x)^2\rmd x$.  For a large negative $q_0$, where localized patterns may be well distinguished (see \fref{fig2}), the term $|q_0|\int(\theta_x)^2\rmd x$ dominates over $\int(\theta_{xx})^2\rmd x$, and the likelihood to observe a non-decaying pattern around $x_0$ can be approximately assessed as the likelihood of $\int q(x)(\theta_x)^2\rmd x$ being positive.

Since $(\theta_x)^2$ varies from pattern to pattern, we can suggest only a qualitative criterion:
\[
q_l(x)\equiv\frac{1}{l}\int_{x-l/2}^{x+l/2}q(x_1)\,\mathrm{d}x_1\,,
\]
where $l$ is the reference size of pattern.  The function $q_l(x)$ is also a convenient representative for illustration of realizations of $q_0+\xi(x)$, because $\delta$-correlated noise $\xi(x)$ possesses infinite mean-square value and cannot be plotted.


\begin{figure}[!t]
\center{
  \includegraphics[width=0.52\textwidth]%
 {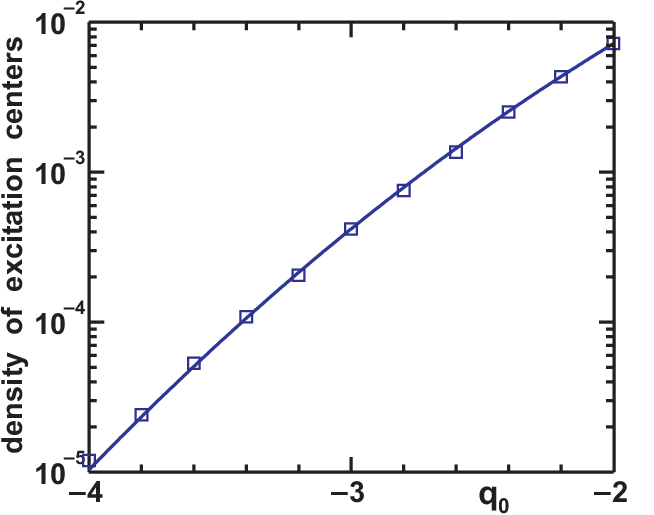}
}
  \caption{Spatial density of centers of convective current excitation in system~\eref{eq2-01} for $u=0$. Squares: results of numerical simulation, solid line: dependence~\eref{rho_app}.}
  \label{fig3}
\end{figure}

We calculated the spatial density of excitation centers of localized patterns as a function of $q_0$. Technical details of numerical simulations can be found in the Appendix. The dependence is plotted in~\fref{fig3}. Remarkably, the calculated dependence can be fairly well fitted with the empirically inferred dependence
\begin{equation}
\rho\approx0.25\,P(q_{l=1.95}>0)
 \approx\frac{\exp(-q_0^2l/4)}{4(-q_0)\sqrt{\pi l\,}}\,.
\label{rho_app}
\end{equation}
Here the probability $P(q_l>0)$ of $q_l$ being positive at given point is the error function of $(q_0\sqrt{l}/2)$. \Eref{rho_app} with $l/2\approx1$ suggests that the above assessments and the underlying interpretation of the occurrence of localized patterns are seminal.  Notice, the notion of quantity $\rho$ makes sense only when excitation centers are separable, {\it i.e.}, the characteristic inter-center distance is large compared to the localization length. Patterns in~\fref{fig2} illustrate that these centers are well separable for $q_0\lesssim-1.5$.

\section{Localization exponents}\label{sec4}

\subsection{Spatial Lyapunov exponents}\label{sec41}
In \sref{sec22}, we have already explained that the localization exponents of the patterns arising in system~\eref{eq2-01} are determined by equation~\eref{eq2-02}. \Eref{eq2-02} with $S=0$ may be rewritten in the form of stochastic system;
\begin{equation}
\theta'=\psi,\quad
\psi'=\phi,\quad
\phi'=-[q_0+\xi(x)]\psi-u\theta\,.
\label{eq31-01}
\end{equation}
One may treat the system~\eref{eq31-01} as a dynamic one with the spatial coordinate $x$ instead of time and evaluate (spatial) Lyapunov exponents (LE), which yield eventually the localization exponents ({\it e.g.}, in review~\cite{Gredeskul-Kivshar-1992} the LE is employed as a localization exponent in order to estimate the localization length in classical AL).

\begin{figure}[!t]
\center{\sf
(a)\hspace{-5mm}\includegraphics[width=0.42\textwidth]%
 {goldobin-shklyaeva-fig04a.eps}\qquad
(b)\hspace{-5mm}\includegraphics[width=0.50\textwidth]%
 {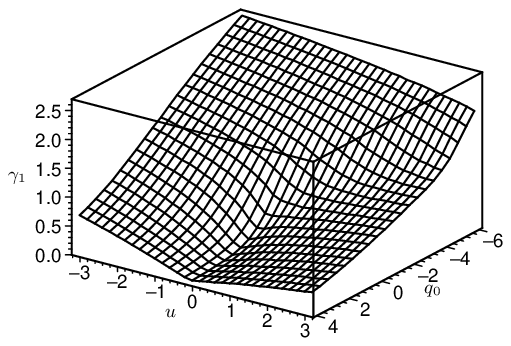}\\[12pt]
(c)\hspace{-5mm}\includegraphics[width=0.50\textwidth]%
 {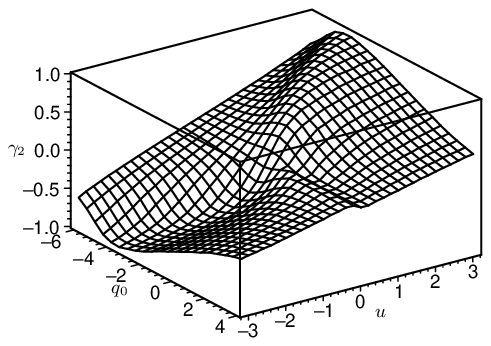}
}
  \caption{Spectrum of Lyapunov exponents is numerically evaluated from runs of system~\eref{eq31-01}. Spectrum $\gamma_1\ge\gamma_2\ge\gamma_3$ at $q_0=-1$ demonstrates the spectral properties in plot~(a). The dependencies of $\gamma_1$ and $\gamma_2$ on $u$ and $q_0$ are presented in plots (b) and (c), respectively.}
  \label{fig4}
\end{figure}

The spectrum of LEs consists of three elements:
\[
\gamma_1\ge\gamma_2\ge\gamma_3\,.
\]
As $\xi(x)$ possesses spatially uniform and isotropic statistical properties and an even distribution, system~\eref{eq31-01} is statistically invariant with respect to the transformation
\[
(u,x,\theta,\psi,\phi)\to(-u,-x,\theta,-\psi,\phi)\,.
\]
When this transformation changes $u$ to $-u$, it simultaneously turns $\gamma_1(u)$, $\gamma_2(u)$, $\gamma_3(u)$ into $-\gamma_1(u)$, $-\gamma_2(u)$, $-\gamma_3(u)$. After arrangement in descending order, one finds $-\gamma_3(u)\ge-\gamma_2(u)\ge-\gamma_1(u)$, which should be the same spectrum $\gamma_1(-u)\ge\gamma_2(-u)\ge\gamma_3(-u)$. Therefore,
\begin{eqnarray}
&\gamma_1(q_0,u)=-\gamma_3(q_0,-u)\,,&\label{eq31-02}\\[5pt]
&\gamma_2(q_0,u)=-\gamma_2(q_0,-u)\,.&\label{eq31-03}
\end{eqnarray}
As the divergence of the phase flow of system~\eref{eq31-01} is zero, $\gamma_1+\gamma_2+\gamma_3=0$; therefore,
\begin{equation}
\gamma_2(q_0,u)=-\gamma_1(q_0,u)+\gamma_1(q_0,-u)\,.
\label{eq31-04}
\end{equation}
The properties~\eref{eq31-02}--\eref{eq31-03} are demonstrated in \fref{fig4}a with the spectrum of LEs for $q_0=-1$. Thus, due to~\eref{eq31-02} and \ref{eq31-04}, it is enough to calculate the largest LE $\gamma_1$ as a function of $u$.

Notice, relationship~\eref{eq31-03} requires $\gamma_2(q_0,u=0)=0$. For $u=0$ the system admits the homogeneous solution $\{\theta,\psi,\phi\}=\{1,0,0\}$, which corresponds to the very LE $\gamma_2=0$.

\Fref{fig4} shows the dependence of $\gamma_1$ (b) and $\gamma_2$ (c) on advection velocity $u$ and mean supercriticality $q_0$. For any $u$, the decrease of $q_0$ leads to the growth of $\gamma_1$, {\it i.e.}, makes the localization more pronounced. For $q_0<0$, where localized flow patterns may occur, and non-large $u$, $\gamma_1$ decreases as $u$ increases. In the following we will also disclose the importance of the fact, that $\gamma_2$ is positive for positive $u$.

Now we come to discussion of relationships between the LEs evaluated and the solution properties. Let us consider some pattern localized near $x_0$ for $u=S=0$. At the distance from $x_0$, on the left side, the solution is a superposition of all the eigenmodes not growing as $x$ tends to $-\infty$, {\it i.e.} the
ones corresponding to LEs ${\gamma_1>0}$ and ${\gamma_2=0}$ (the eigenmode of the latter is spatially homogeneous; $\Theta_2=const$): $\Theta_{2,-}+\Theta_1(x)\,e^{\gamma_1(x-x_0)}$. Here $\Theta_1(x)$ is a bounded function, which neither decays nor grows averagely over large distances, determined by a specific realization of noise. On the right side, the solution is, similarly, a superposition the eigenmodes corresponding to LEs ${\gamma_3<0}$ and ${\gamma_2=0}$: $\Theta_{2,+}+\Theta_3(x)\,e^{\gamma_3(x-x_0)}$, where $\Theta_3(x)$ is a bounded function as well as $\Theta_1(x)$. Thus, at the distance from $x_0$, one can write
\[
\theta(x)\approx\left\{\begin{array}{cl}
\Theta_{2,-}+\Theta_1(x)\,e^{\gamma_1(x-x_0)},&
\mbox{ for }x<x_0,\\[5pt]
\Theta_{2,+}+\Theta_3(x)\,e^{\gamma_3(x-x_0)},&
\mbox{ for }x>x_0.
\end{array}\right.
\]
Indeed, in~\fref{fig6}a one can see the temperature profile to tend exponentially to different constant values $\Theta_{2,-}$ and $\Theta_{2,+}$ on the left and right sides of the excitation area. Considering the solution in between of two excitation domains near $x_1$ and $x_2>x_1$, one can combine the above asymptotic laws for $x<x_0$ (assuming $x_0=x_2$), {\it i.e.}, $\Theta_{2,-}+\Theta_1(x)\,e^{\gamma_1(x-x_2)}$, and for $x>x_0$ (assuming $x_0=x_1$), {\it i.e.}, $\Theta_{2,+}+\Theta_3(x)\,e^{\gamma_3(x-x_1)}$, and obtain
\begin{equation}
\theta(x_1<x<x_2)\approx\Theta_3(x)e^{\gamma_3(x-x_1)}+\Theta_2
+\Theta_1(x)e^{\gamma_1(x-x_2)}.
\label{eq31-05}
\end{equation}
Here the amplitude of the corresponding stream function
\[
\psi(x)\approx\Psi_3(x)e^{\gamma_3(x-x_1)}+\Psi_1(x)e^{\gamma_1(x-x_2)},
\]
where $\Psi_{1,3}(x)$ are bounded similarly to $\Theta_{1,3}(x)$, is localized near $x_1$ and $x_2$ with the exponent $\gamma_1$ (for $u=0$, $\gamma_3=-\gamma_1$). Meanwhile, the temperature perturbation~\eref{eq31-05} is not localized because of constant $\Theta_2$ which is generally different between different neighboring excitation areas (\fref{fig7}a).

Let us consider the case of $u>0$ (the case of $u<0$ is similar and does not require special discussion).
For $u>0$, the shift of the temperature $\theta\to\theta+S/u$ eliminates the heat flux $S$. From the claim $S=0$ for $u\ne0$, it follows that in the domain of the flow damping, where $\psi\to0$, the temperature perturbation $\theta$ tends to zero as well. Indeed, solution~(\ref{eq31-05}) takes the form
\begin{equation}
 \theta(x_1<x<x_2)\approx\Theta_3(x)\,e^{\gamma_3(x-x_1)}
 +\Theta_2(x)\,e^{\gamma_2(x-x_2)}
   +\Theta_1(x)\,e^{\gamma_1(x-x_2)};
\label{eq31-06}
\end{equation}
here $\gamma_2>0$, and therefore the mode corresponding to $\gamma_2$ is localized near $x_2$. For $u\ne0$, the contribution of the second term of equation~\eref{eq31-06} tends to $0$ at the distance from the excitation domains. Hence, in the presence of the advection the temperature perturbations are localized as well as the fluid currents (for example, see \fref{fig8}).

On the other hand, now the mode corresponding to $\gamma_2$ makes a nonzero contribution to the flow:
\[
\psi(x)\approx\Psi_3(x)e^{\gamma_3(x-x_1)}
+\Psi_2(x)e^{\gamma_2(x-x_2)} +\Psi_1(x)e^{\gamma_1(x-x_2)},
\]
illustration of which can be found in \fref{fig8}. For $\gamma_2>0$, the flow is localized down the advection stream (the right flank of the localized pattern) with the exponent $\gamma_3$; $\gamma_-=|\gamma_3|$. On the left flank of the excitation domain, two modes appear:
\begin{equation}
\psi(x<x_2)\approx\Psi_2(x)\,e^{\gamma_2(x-x_2)}
+\Psi_1(x)\,e^{\gamma_1(x-x_2)}.
\label{eq31-07}
\end{equation}
For moderate $u$, $\Psi_1(x)$ and $\Psi_2(x)$ are comparable; hence, the mode $\Psi_1(x)e^{\gamma_1(x-x_2)}$ rapidly ``disappears'' against the background of $\Psi_2(x)e^{\gamma_2(x-x_2)}$ as one moves away from $x_2$, and the upstream localization properties are determined by the persisting $\gamma_2$-mode; $\gamma_+=\gamma_2$. For $u=0$, the function $\Psi_2(x)=0$, and, for small $u$, $\Psi_2(x)$ remains small by continuity. Hence, for vanishingly small $u$ the flow~\eref{eq31-07} considerably decays in the domain where the $\gamma_1$-mode remains dominating over the small $\gamma_2$-one, and this mode determines the upstream localization length; $\gamma_+=\gamma_1$.

\subsection{Growth exponents of mean-square values}\label{sec42}
For an analytical estimation of the largest LE of a linear stochastic system, one may calculate the exponential growth rate of mean-square values of variables ({\it e.g.}, see~\cite{Klyatskin-2005}). Specifically, we employ the following particular result of~\cite{Klyatskin-2005}, which is valid for a linear system of ordinary differential equations with noisy coefficients;
\[
y'_i=\sum_jL_{ij}\,y_j+\xi(x)\sum_j\Gamma_{ij}\,y_j\,.
\]
For normalized Gaussian $\delta$-correlated noise $\xi(x)$ (\,$\left\langle\xi(x)\right\rangle=0$, $\left\langle\xi(x+x')\xi(x)\right\rangle=2\delta(x')$\,), the mean values $\langle y_i\rangle$ (averaged over noise realizations) obey the equation system
\begin{equation}
\langle y_i\rangle'
 =\sum_j(\mathrm{\bf L}+{\bf\Gamma}^2)_{ij}\langle y_j\rangle\,.
\label{kljaz}
\end{equation}
(Klyatskin derived this result for $\delta$-correlated noise which is not necessarily Gaussian. A simple rederivation of this formula for Gaussian noise can be found in~\cite{Zillmer-Pikovsky-2005}, where the exponential growth rate of mean-square values was utilized for an approximate analytical calculation of the Lyapunov exponent for a stochastic system similar to system~\eref{eq31-01}).

\begin{figure}[!t]
\center{\sf
(a)\hspace{-10mm}\includegraphics[width=0.52\textwidth]%
 {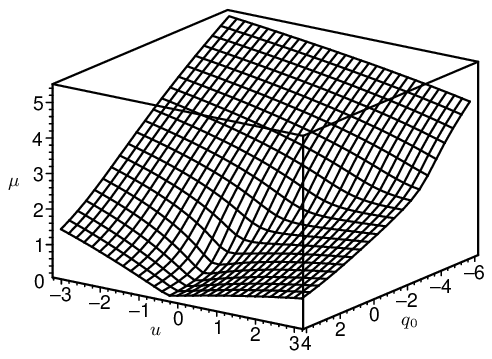}\\[20pt]
(b)\hspace{-5mm}\includegraphics[width=0.62\textwidth]%
 {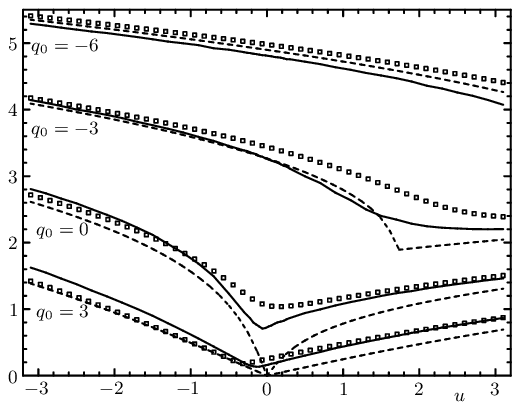}
}
  \caption{(a):~Largest growth exponent $\mu$ of the mean-square temperature is plotted for system~\eref{eq31-01}. (b):~Exponent $\mu$ (squares) is compared to doubled largest Lyapunov exponents $2\gamma_1$ of system~\eref{eq31-01} (solid line) and of the noiseless system (dashed line).
}
  \label{fig5}
\end{figure}

For $\vec{y}=\{\theta,\psi,\phi\}$, matrix ${\bf\Gamma}^2=0$, and noise does not manifest itself. This is due to the system symmetry and does not characterize behavior of a particular system under given realization of noise. In order to characterize this behavior, one has to consider behavior of mean-square values (cf.\,\cite{Klyatskin-2005}). For $\vec{y}=\{\theta^2,\theta\psi,\theta\phi,\psi^2,\psi\phi,\phi^2\}$, equations~\eref{eq31-01} yield
\begin{eqnarray}
&&y_1'=2y_2\,,\nonumber\\
&&y_2'=y_4+y_3\,,\nonumber\\
&&y_3'=y_5-[q_0+\xi(x)]y_2-uy_1\,,\nonumber\\
&&y_4'=2y_5\,,\nonumber\\
&&y_5'=y_6-[q_0+\xi(x)]y_4-uy_2\,,\nonumber\\
&&y_6'=-2[q_0+\xi(x)]y_5-2uy_3\,.\nonumber
\end{eqnarray}
Hence,
\[
 \mathrm{\bf A}\equiv\mathrm{\bf L}+{\bf\Gamma}^2=\left[\begin{array}{cccccc}
 0&2&0&0&0&0\\
 0&0&1&1&0&0\\
 -u&-q_0&0&0&1&0\\
 0&0&0&0&2&0\\
 0&-u&0&-q_0&0&1\\
 0&0&-2u&2&-2q_0&0
\end{array}\right].
\]

The growth exponent $\mu$ of mean-square values is the largest real (as a mean-square value cannot oscillate) meaningful eigenvalue of matrix $\mathrm{\bf A}$ (\footnote{A formal eigenvector is meaningful, if it meets the conditions
 $\langle\theta^2\rangle\ge0$, $\langle\psi^2\rangle\ge0$,
 $\langle\phi^2\rangle\ge0$,
 $\langle\theta^2\rangle\langle\psi^2\rangle\ge\langle\theta\psi\rangle^2$,
 $\langle\theta^2\rangle\langle\phi^2\rangle\ge\langle\theta\phi\rangle^2$,
 and $\langle\psi^2\rangle\langle\phi^2\rangle\ge\langle\psi\phi\rangle^2$.}).
Although the characteristic equation of matrix $\mathrm{\bf A}$ is a 6-degree polynomial of $\mu$, it is just a quadratic polynomial of $u$, and one can analytically find the surface of the largest real $\mu(q_0,u)$ in the parameter space in a parametric form:
\begin{eqnarray}
&&
 u_{1,2}=\frac{7}{16}\mu^3+\frac{1}{4}q_0\mu-\frac{1}{2}
\nonumber\\[3pt]
&&\qquad
 \pm\sqrt{\frac{81}{256}\mu^6\!+\!\frac{27}{32}q_0\mu^4
 \!-\!\frac{15}{16}\mu^3\!+\!\frac{9}{16}q_0^2\mu^2\!-\!\frac{3}{4}q_0\mu\!+\!\frac{1}{4}}\,.
\label{eq32-01}
\end{eqnarray}
These expressions make sense for $\mu>0$, $q_0\ge2/3\mu-3\mu^2/4+\sqrt{2\mu/3}$. \Fref{fig5} presents
this surface and demonstrates good agreement between $2\gamma_1$ and $\mu$ (which do not necessarily coincide by definition).

\begin{figure}[!t]
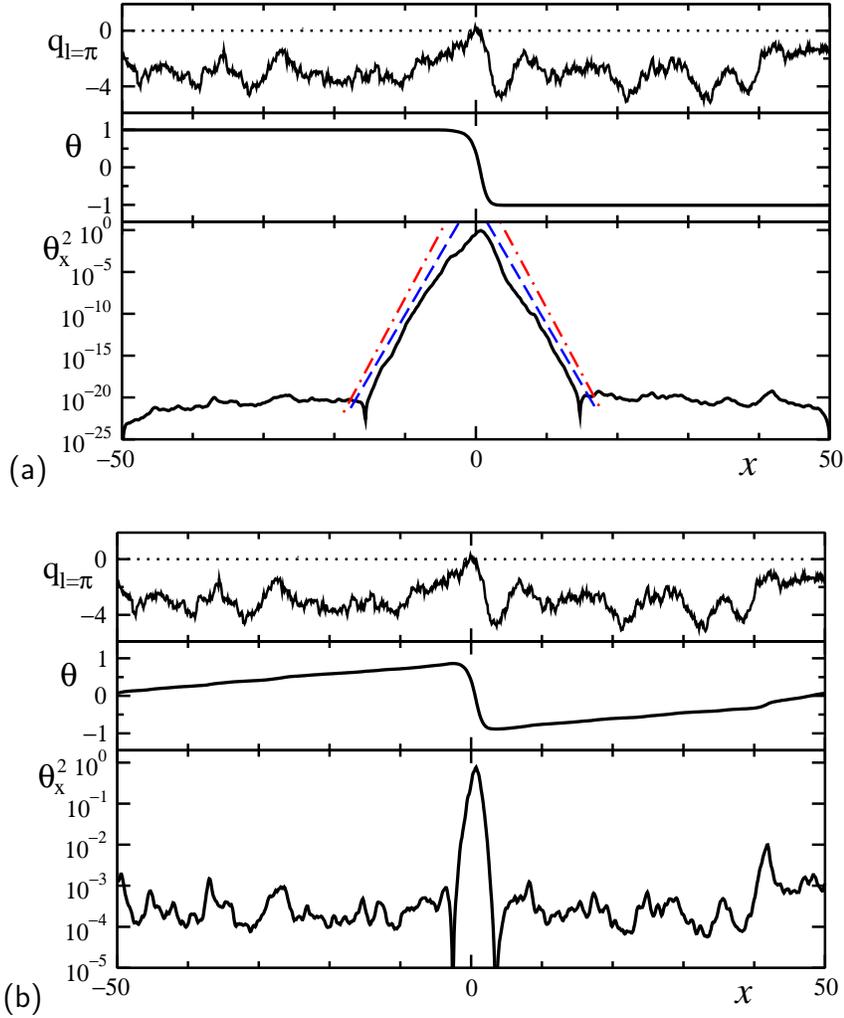

\center{\sf
(a)\includegraphics[width=0.68\textwidth]%
 {goldobin-shklyaeva-fig06a.eps}\\[15pt]
(b)\includegraphics[width=0.68\textwidth]%
 {goldobin-shklyaeva-fig06b.eps}
}
  \caption{
Sample steady time-independent solutions to equation~\eref{eq2-01} without imposed advection for $q_0=-3.1$ (realization of $q(x)$ are represented by $q_{l=\pi}(x)$). The lateral boundaries of the domain ($x=\pm50$) are thermally-insulating impermeable in (a) and periodic in (b). Dashed lines: the inclination corresponding to flow decay with the exponent $\gamma_1=1.67$, dashdot lines: decay with the exponent $\mu/2=1.81$.}
  \label{fig6}
\end{figure}

\begin{figure}[!t]
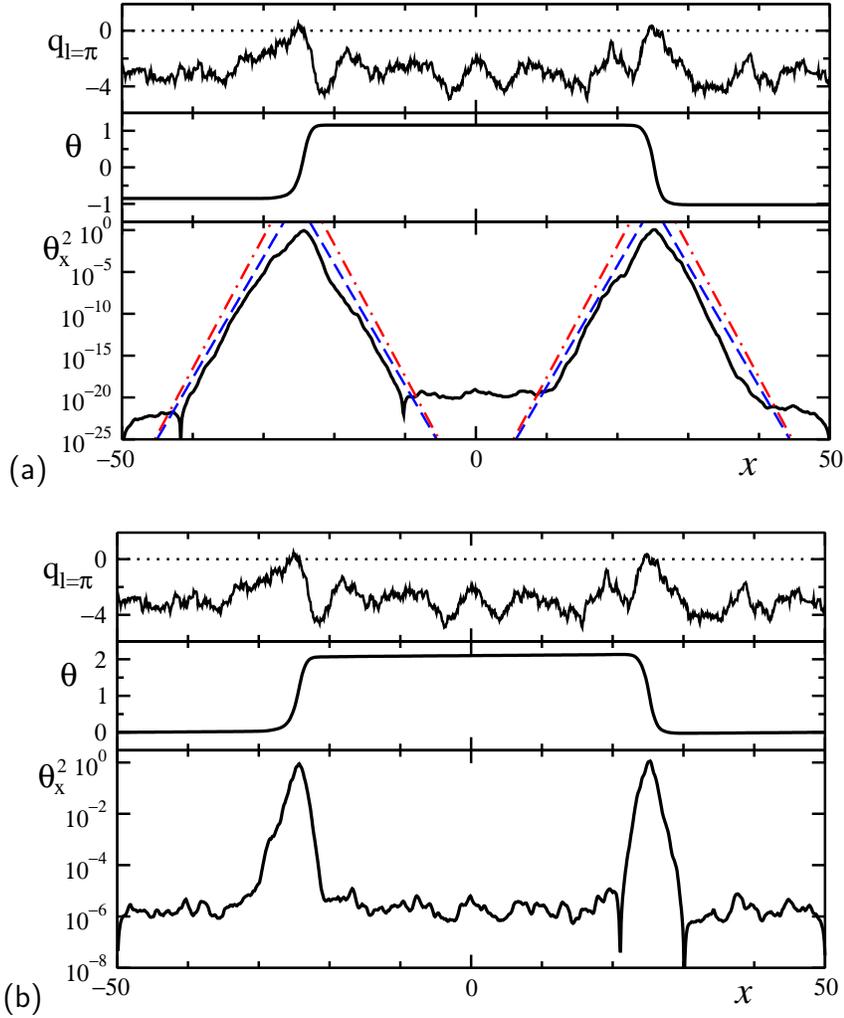

\center{\sf
(a)\includegraphics[width=0.68\textwidth]%
 {goldobin-shklyaeva-fig07a.eps}\\[15pt]
(b)\includegraphics[width=0.68\textwidth]%
 {goldobin-shklyaeva-fig07b.eps}
}
  \caption{
Sample steady time-independent solutions to equation~\eref{eq2-01} with two nearby localized patterns ($u=0$, $q_0=-3.1$). The lateral boundaries of the domain are thermally-insulating impermeable in (a) and isothermal impermeable in (b). Dashed lines: the inclination corresponding to flow decay with the exponent $\gamma_1=1.67$, dashdot lines: decay with the exponent $\mu/2=1.81$.}
  \label{fig7}
\end{figure}

\begin{figure}[!t]
\center{\includegraphics[width=0.68\textwidth]%
{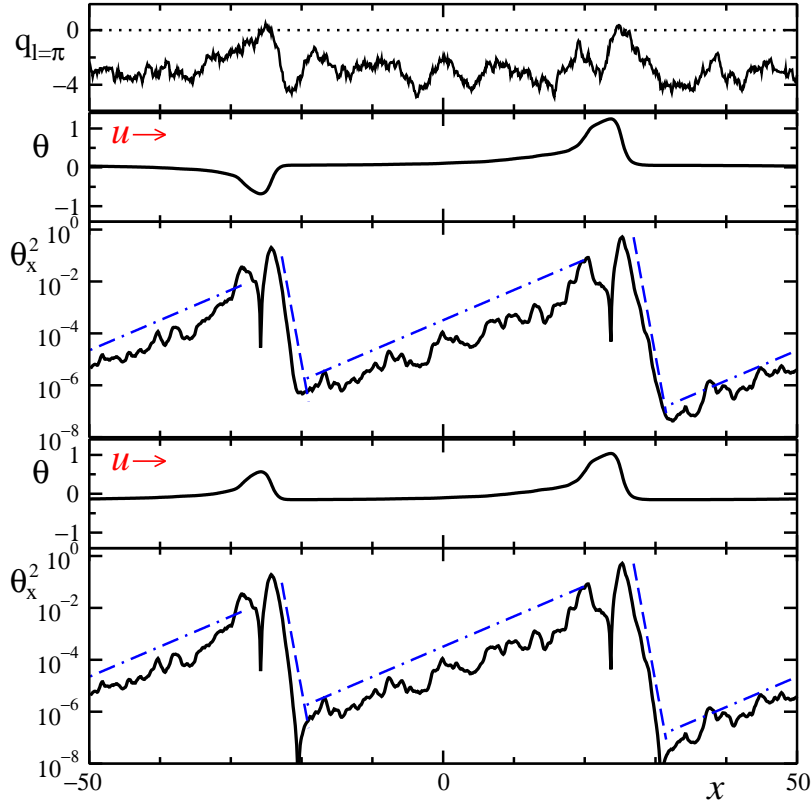}}
  \caption{
Sample steady time-independent solutions to equation~\eref{eq2-01} for $u=0.3$, $q_0=-3.0$ and periodic lateral boundaries. Dashed lines: the inclination corresponding to flow decay with the exponent $\gamma_3=1.70$, dashdot lines: decay with the exponent $\gamma_2=0.134$. One also observes multistability ({\it i.e.}, coexistence of states stable to small perturbations) for each separate localized flow pattern; it can change its sign.}
  \label{fig8}
\end{figure}

\section{Nonlinear solutions}\label{sec5}
In previous sections we have analysed the spatial density of localized patterns excited in the system and studied the localization properties of patterns given these patterns are excited. The reported results on spatial density and localization are statistical and can be derived from the linearization of equation~\eref{eq2-01} in general. Let us now consider actual nonlinear patterns arising in the system.

Specifically, in this section our tasks are:
\\
(i)\;to underpin the findings of \sref{sec4} with the results of numerical simulation for the nonlinear problem~\eref{eq2-01},
\\
(ii)\;to explore the role of lateral boundary conditions for a finite layer, and
\\
(iii)\;to discuss the consequences of the fact that the flow patterns may be localized up the advection stream with exponent $\gamma_2$ which can be small.
\\
This will serve our ultimate goal: to check and to show, that all the effects and localization features found with the linear system can be observed in the full nonlinear problem. We will also have validation of the argument of \sref{sec22}, that the localization properties can be comprehensively derived from equation~\eref{eq2-02}.

First, the behavior of the fluid dynamical system~\eref{eq2-01} in the absence of imposed advection requires consideration. In figures~\ref{fig6} and \ref{fig7}, one can see samples of the localized patterns arising in the system. With thermally insulating lateral boundaries (\fref{fig6}a), the localized flow is excited in the vicinity of the domain, where $q_{l=\pi}$ spontaneously takes positive values, and decays beyond the excitation domain (the background noise $\theta_x^2\sim10^{-25}$ is due to the limitation of numeric
accuracy). Thermally insulating lateral boundary conditions for a finite observation domain are eventually ``free'' ones (no external impositions at the boundary) and thus correspond to a large (compared to the domain size) distance between excitation centers in an infinite layer. For $q_0=-3.1$, this distance is indeed large, approximately $3\cdot10^3$ (see \fref{fig3}).

For periodic lateral boundary conditions (\fref{fig6}b), the finiteness of the distance between excitation centers effectively manifests itself (this distance is the system period). The system~\eref{eq2-02} with vanishing noise admits the trivial solution $\theta=q_0^{-1}S\,x$ which is related to a constant nonzero heat flux $S$. In the presence of noise, this mode turns into the time-independent solution $\theta'=q_0^{-1}S+\psi_1$, $\left\langle\psi_1\right\rangle=0$, which, for small $S$, obeys the equation
\[
\psi_1''+[q_0+\xi(x)]\psi_1=-q_0^{-1}S\,\xi(x)\,,
\]
(here $\langle\xi(x)\psi_1\rangle=0$), {\it i.e.}, $\psi_1\propto S$. In the presence of noise this solution dominates in the region where flows are damped ($q_l<0$). In the situation presented in \fref{fig6}b, $q_0^{-1}S=[\theta]/L_\mathrm{layer}$, where $[\theta]\approx2$ is the temperature altitude imposed by the excited nonlinear flow, $L_\mathrm{layer}\approx100$ is the system period (calculation domain size). The quantity $(S/q_0)^2\approx4\cdot10^{-4}$ has exactly the order of magnitude corresponding to the background flow observed at the distance from the excitation center in \fref{fig6}b. This background flow distorts the localization and makes the localization exponents of the ``tails'' of the excited localized pattern hardly measurable. For a time-independent pattern the heat flux $S$ is constant over the layer; therefore the mean background flow $\langle\Psi\rangle\propto\langle\theta'\rangle=q_0^{-1}S$ remains constant over the layer even for a large number of excitation centers.

For thermally insulating lateral boundaries the heat flux $S$ related to the background flow should decay to zero. Therefore with thermally insulating lateral boundaries the time-independent localized patterns occur without the background flow for an arbitrary number of excitation centers (\fref{fig7}a).

With isothermal lateral boundaries (``isothermal'' means that they are maintained under constant temperature $\theta=0$) the intensity of the background flow is nonzero, though it decreases as the layer extends. The cause of the decrease is that, for practically separated localized patterns, there is multistability between solutions with different signs of the temperature jumps in domains of intense currents. Hence, with a large
number of excitation centers, the jumps can be mutually balanced, resulting in a small value of $q_0^{-1}S=L_\mathrm{layer}^{-1}\sum_j[\theta]_j$; {\it e.g.}, in \fref{fig7}b the background flow is considerably smaller than in \fref{fig6}b. Strictly speaking, for independent random $[\theta]_j$, the sum $\sum_j[\theta]_j\propto\sqrt{L_\mathrm{layer}}$, and, hence, $q_0^{-1}S\propto L_\mathrm{layer}^{-1/2}$ tends to zero as $L_\mathrm{layer}$ tends to infinity.

Sample flows for $u\ne0$ are shown in \fref{fig8}. In agreement with predictions of \sref{sec41}, the temperature is equal on the both sides of an intense flow domain, {\it i.e.}, not only flows are localized but also temperature perturbations. And, which is more noteworthy, even for quite small $u$ when $\gamma_2$ is also small, the localization properties up the imposed advection stream are determined by this small $\gamma_2$, but not by $\gamma_1$ which is in force for vanishing $u$ as explained in \sref{sec41}. Small localization exponent means a large localization length, which is a drastic change to the localization properties. For instance, in \fref{fig8} for $u=0.3$ the upstream localization length is increased by factor 12 compared to the one in the absence of the imposed advection.

Noteworthy, the advectionally increased localization length becomes comparable to the mean distance between the excitation centers (see \fref{fig3}) for moderate negative $q_0$. In this way, weak advection may lead to the transition from a set of localized convective flow areas to an almost everywhere intense ``global'' flow, and, for instance, drastically enhance transport of a nearly indiffusive pollutant which is transferred rather convectively than by the molecular diffusion. Quantitative analysis of these effects is a separate and laborious physical problem and will be considered elsewhere (this problem has already been addressed in~\cite{Goldobin-Shklyaeva-2009-Goldobin-2010}).

\section{Comparison with the Anderson localization in conservative systems}\label{sec6}
The problem we consider is different from the one of the Anderson localization in the Schr\"odinger equation not only due to the imposed advection $u$, but also (and in the first place) in the physical interpretation and observability of effects related to properties of formal solutions.

In the linear Schr\"odinger equation, different localized solutions do not mutually interact. In our fluid dynamical system, all these modes do mutually interact via nonlinearity and build up into a net stationary flow. This net flow may be almost everywhere intense given the spatial density of localized modes is high enough, and the identity of separate localized contributions can be completely lost because of nonlinearity. The localization properties of self-excited patterns become important only when they are rare in space and well separated from each other.

In quantum mechanics, the nonlinear Schr\"odinger equation appears as a model reduction in the many-body problem. There, the nonlinearity was also reported to lead to destruction of AL ({\it e.g.}, \cite{Gredeskul-Kivshar-1992,Pikovsky-Shepelyansky-2008,Mulansky-Pikovsky-2013} and \cite{Mulansky-Pikovsky-2012a} for two-dimensional lattices). Moreover, it has been revealed to be never ``weak'' in this relation; the smaller the nonlinearity the larger decay time of a localized mode is, but the localized mode does not persist infinitely long for arbitrary small nonlinearity. It was also discovered that the finiteness of the response time ({\it i.e.}, the presence of relaxation in the nonlinear terms) turns delocalization into re-localization~\cite{Caetano-Moura-Lyra-2011,Mulansky-Pikovsky-2012b}, subdiffusive spreading of an initially localized excitation is suppressed. Nonetheless, even when the localization is imperfect, the localized modes still have a physical meaning in quantum mechanics (or in acoustics). Additionally, the physical meaning of the quantum wave function imposes strong limitations on the form of nonlinearity (the particle conservation law, {\it etc.}), whereas in a fluid dynamical system similar limitations are not applied.

To summarize, firstly, the results of studies for nonlinear effects similar to~\cite{Gredeskul-Kivshar-1992,Pikovsky-Shepelyansky-2008,Mulansky-Pikovsky-2013,Mulansky-Pikovsky-2012a} may not (not always) be directly applicable to the case of dissipative systems in fluid dynamics. Secondly, the localization of linear modes itself is not always of significance for the system (it may be completely insignificant for the above-mentioned case of a high spatial density of localized modes, where they are ``lost'' in an intense nonlinear net flow). In other words, the localization problem in dissipative systems is relevant only when the self-excited patterns are rare enough.

\section{Conclusion}\label{concl}
We have studied the localization phenomenon in the problem of thermal convection in a thin horizontal layer subject to random spatial inhomogeneity and considered the effect of an imposed longitudinal advection on localization properties. The study relies on the equation~\eref{eq2-01} which is relevant to a broad variety of fluid dynamical systems~\cite{Knobloch-1990,Shtilman-Sivashinsky-1991,Aristov-Frick-1989,Goldobin-Shklyaeva-BR-2008} and some other dissipative media where pattern formation occurs. For instance, the Kuramoto--Sivashinsky equation ({\it e.g.}, see~\cite{Michelson-1986} and refs.\ therein for examples of physical systems) has the same linear part as equation~\eref{eq2-01} with $u=0$. Moreover, the basic laws in physics are conservation ones, which quite often remains reflected in the final equations having the form
 $\partial_t\mbox{[quantity]}+\nabla\cdot\mbox{[flux of quantity]}=0$;
{\it e.g.}, in~\cite{Hoyle-1998} the equation for the Eckhaus instability mode in a system relevant to binary convection at small Lewis number~\cite{Schoepf-Zimmermann-1989-1993} preserves such a form. With such conservation laws either for systems with the sign inversion symmetry of the fields, which is quite typical in physics, or for description of spatial modulation of an oscillatory mode, the Kuramoto--Sivashinsky equation should be modified exactly to equation~\eref{eq2-01}. For thermal convection in a porous layer~\cite{Goldobin-Shklyaeva-BR-2008}, the frozen parametric disorder $\xi(x)$ may be due to random inhomogeneities of the porous matrix (which are inevitable in real world), while the mean supercriticality $q_0$ may be controlled in experiments. Thus, our study can be expected to be not only far from being related to something artificial in thermal convection, but also rather general than specific to the thermal convection in porous media.

Summarizing, localized nonlinear flow patterns have been observed below the instability threshold of the system without disorder, and the dependence of the spatial density of the localized current patterns on the mean deviation $q_0$ from this threshold has been found numerically (see \fref{fig3} and approximate expression~\eref{rho_app}). The up- and downstream localization exponents have been evaluated numerically (\fref{fig4}) and estimated analytically (equation~\eref{eq32-01}). In particular, the imposed advection has been found to result in the localization of the temperature patterns in addition to the convective currents (the former are not localized in the absence of advection). In agreement with theoretical predictions, numerical simulation for the nonlinear equation has exhibited a crucial effect of the imposed advection on the upstream localization properties; the localization length can increase by one order of magnitude for small finite $u$. Via the upstream delocalization, weak advection may lead to the transition from a set of localized current patterns to an almost everywhere intense ``global'' flow, and, {\it e.g.}, essentially enhance transport of a nearly indiffusive pollutant~\cite{Goldobin-Shklyaeva-2009-Goldobin-2010}.

\section*{Acknowledgements}
Authors are grateful to Dmitry Lyubimov for interesting and seminal discussions and comments and acknowledge financial support by Grant of The President of Russian Federation (MK-6932.2012.1) and by the Government of Perm Region (Contract C-26/212).

\appendix
\setcounter{section}{1}
\section*{Appendix: Numerical simulations with equation~\eref{eq2-01}}\label{appendix}
All the numerical simulations for equation~\eref{eq2-01} were performed by means of the finite difference method.  The specification of the discretization scheme is as follows: the first $x$-derivatives are central, the noiseless $x$-derivatives have accuracy $(\rmd x)^2$, the time step $\rmd t=(\rmd x)^4/11$, the noise $\xi(x)$ is generated in the middle between the mesh nodes. The space step $\mathrm{d}x=0.05$ appeared to be small enough for calculations which is ensured (i)\,by the results in figures \ref{fig6} and \ref{fig8}, where the localization properties of calculated patterns are in agreement with the independent and much more accurate simulation for null-dimensional system~\eref{eq31-01} and (ii)\,by the fact, that statistical properties do not change when step $\rmd x$ is halved.

As the localized patterns are rare, especially for large negative $q_0$, a non-trivial approach to the simulation was required.  The procedure of calculations for localized patterns was as follows:
\\
(1)\ Realization $q(x)$ was generated for an interval of length $10000$, and corresponding $q_{l=\pi}(x)$ was calculated for the entire interval.
\\
(2)\ Local extrema of $q_{l=\pi}(x)$ were indexed and indexes were arranged in the order of decrease of the extreme value.
\\
(3)\ Domain of length $2H$ with the extreme value of $q_{l=\pi}(x)$ was cut out from the large interval and equation~\eref{eq2-01} was simulated on this shortened interval. We did not set $2H$ lesser than $50$. Notice, for step~(2) the $2H$-interval around some extremum of $q_{l=\pi}$ was excluded from consideration for smaller extrema; the smaller extrema in this subinterval were not indexed.

This procedure allows simulation of localized patterns even when they are extremely rare, and does not affect the statistical properties of $q(x)$ in the simulation domain.

For calculations of the spatial density of the excitation centers of localized patterns, we successively performed simulation for subdomains with extrema.  If the temperature field $\theta(x)$ did not decay to a uniform state, the domain was counted as the one with pattern excited.  The criterion to finish calculations for smaller extrema was having $N+\sqrt{M}$ domains with no excitation in sequence, were $M$ is the number of already detected excitation domains and $N=30$.  Tests revealed that the results do not change for $N$ increasing beyond $10$, meaning $N=30$ is a reliable choice for criterion.

~\

\bibliographystyle{elsarticle-num}

\end{document}